\documentclass{article}

\newfont{\bbb}{msbm10 scaled 500}

\newfont{\bb}{msbm10 scaled 1100}

\newcommand{\lv}{{\bf l}}

\newcommand{\sv}{{\bf s}}

\newcommand{\yv}{{\bf y}}
\newcommand{\zv}{{\bf z}}

\newcommand{\Hm}{{\bf H}}

\renewcommand{\det}{{\hbox{det}}}

\usepackage{eusipco2011}       
\usepackage{graphicx}    
\usepackage{amsfonts}
\usepackage{amssymb}
\usepackage{amsmath}
\usepackage[hang,small,bf]{caption}                   
\DeclareGraphicsExtensions{.pdf,.png,.jpg}

\title{Impact of Mobility on MIMO Green Wireless Systems}

\name {Vineeth S Varma$^{1,2}$, Merouane Debbah$^3$, Samson Lasaulce$^2$ and Salah Eddine Elayoubi$^1$}

\address{\begin{minipage}{181pt} \centering $^{1}$Orange Labs\\92130 Issy Les Moulineaux\\France\\Email: {vineeth.svarma,\\salaheddine.elayoubi}@orange-ftgroup.com
\end{minipage}
\begin{minipage}{30pt}     
\end{minipage}
\begin{minipage}{150pt}\centering  $^{2}$LSS, SUPELEC\\ 91192 Gif sur Yvette\\ France\\Email: {Vineeth.VARMA,\\Samson.LASAULCE}@lss.supelec.fr
\end{minipage}
\begin{minipage}{30pt}  \space    
\end{minipage}
\begin{minipage}{160pt}\centering  $^{3}$Alcatel-Lucent Chair, SUPELEC\\91192 Gif sur Yvette\\France\\Email: {merouane.debbah}@supelec.fr
\end{minipage}
}

\begin{document}

\maketitle

\begin{abstract}

This paper studies the impact of mobility on the power consumption of wireless networks. With increasing mobility, we show that the network should dedicate a non negligible fraction of the useful rate to estimate the different degrees of freedom. In order to keep the rate constant, we quantify the increase of power required for several cases of interest. In the case of a point to point MIMO link, we calculate the minimum transmit power required for a target rate and outage probability as a function of the coherence time and the number of antennas. Interestingly, the results show that there is an optimal number of antennas to be used for a given coherence time and power consumption. This provides a lower bound limit on the minimum power required for maintaining a green network.
\end{abstract}

\section{Introduction}

The rapidly increasing demand for higher data rates has to be met by network providers. Arbitrarily increasing the signal to noise ratio (SNR) is not possible due to mainly two reasons. First, the transmit power of the Base Stations (BS) has to be constrained for energy efficiency considerations. Secondly, the transmit power has to be regulated due to electromagnetic restriction issues like in \cite{wifir}. As a consequence, it is of great importance to determine the minimum power a transmitter requires to provide a certain target rate in wireless communication.

A multiple input multiple output (MIMO) channel is a well known and promising technique to increase the performance of a point to point link. In essence, MIMO system consists of multiple transmitting antennas and receiving antennas. If we have perfect channel state information (CSI), then increasing the number of antennas would always increase the rate. However, if the channel has to be estimated through training, time has to be spent for training each degree of freedom. We consider the impact of mobility in a MIMO system. In its full generality, this problem should be treated as a non coherent MIMO channel for which the capacity is hard to determine (open problem). Therefore, we restrict ourselves to a training mechanism which is suboptimal but provides tractable expressions. It was shown in \cite{mimot} that in the context of point to point MIMO, there is a tradeoff between the channel estimate and the training time. A training time that maximizes the achievable rate was found in \cite{mimot}. In this work we study how mobility affects the outage rates in MIMO systems.

Our objective is to study the transmit power for a target data rate, outage probability and coherence time. With an infinite coherence time, it is known that increasing the number of antennas would cause the power consumption to decrease. However, if the coherence time is small, a fraction of the time has to be used for channel training which results in a higher power consumption to transmit at the same rate. In this paper, we tackle the problem of finding the optimum number of antennas and training time for single input single output (SISO), single input multiple output (SIMO), multiple input single output (MISO) and MIMO systems. 

The system model is described in Section 2. The performance metric, schemes under which we calculate the transmit power and the objectives of this paper are detailed in Section 3. The power consumed in MISO, SIMO and MIMO systems are calculated analytically and numerically in Sections 4 and 5. The power thus calculated is studied as a function of the coherence time and the number of antennas. Finally, we draw conclusions on the effect of mobility on MIMO systems.

\section{System model}

Let us consider the input-output relationship of a MIMO system given by:
\begin{equation}
\yv=\sqrt{\frac{\rho}{M}}\Hm\sv + \zv
\label{eq:cmodel}
\end{equation}where $\yv \in \mathbb{C}^N$ is the received signal, the dimension $N$ representing the number of receive antennas. The transmitted signal is $\sv \in \mathbb{C}^M$, where $M$ is the number of transmit antennas.$\Hm \in \mathbb{C}^{M \times N}$ represents the channel connecting the $M$ transmit to the $N$ receive antennas, and $\zv \in \mathbb{C}^N$ represents the additive noise. The matrix $\Hm$, and the vectors $\sv$ and $\zv$ all comprise of i.i.d complex Gaussian entries with mean zero and variance unity, that is $h_{ij},s_{i},z_{j} \sim C\mathcal{N}(0,1)$. The total transmit power is proportional to $\rho$ which is the average SNR at each receive antenna. 

We assume that the channel obeys the simple discrete-time block-fading law, where the channel is constant for some discrete time interval, after which it changes to an independent value that it holds for the next interval \cite{mimo}. The coherence time is essentially determined by the mobility of the user and so if we calculate the dependency of the transmit power with respect to the coherence time, we will establish a relation between the mobility of a user and the corresponding transmit power. We further assume that channel estimation (via training) and data transmission is to be done within this coherence time, after which a new training sequence is done. The time for training in symbols will be given by $t$ and the coherence time in symbols by $T$. The inverse of the fraction of time that is used for data transfer will be denoted by $\zeta=(\log_{2}(e)(1 - \frac{t}{T}) )^{-1}$. (The $\log_{2}(e)$ is for the conversion of natural logarithms to base two). We also denote $\frac{t}{M}$ by $\tau$.

In the training phase, all $M$ transmitting antennas broadcast orthogonal sequences of known pilot symbols of equal power $\rho$ on all antennas. The orthogonality condition imposes $\tau \geq 1$. The receiver estimates the channel $\Hm$, based on the observation of the pilot sequence, as $\hat{\Hm}$ and the error in estimation is given as $\tilde{\Hm} = \Hm - \hat{\Hm}$. The channel estimate normalized to variance one is given by $\bar{\Hm}$. From \cite{mimot} we know that the rate is lowest (worst case) when $\tilde{\Hm}$ is Gaussian and then, the channel model can be rewritten as
\begin{equation}
\yv=\sqrt{\frac{\rho_{eff}}{M}}\bar{\Hm}\sv + \bar{\lv}
\label{eq:cmodel2n}
\end{equation} where $\rho_{eff}$ is given by $\frac{\tau\rho^2}{1+\rho + \rho \tau}$ and $\bar{\lv}$ equal to $\rho\tilde{\Hm}\sv+\zv$ normalized to unit variance. (\ref{eq:cmodel2n}) leads to a lower bound on the mutual information and the achievable rate. Thus, all formulas derived in the following sections give lower bounds on the achievable rate and upper bounds on the outage and transmit power. This was verified to be an effective model in other works as well (See \cite{samson}). The value of $\rho$ can be calculated from $\rho_{eff}$ by inverting the equation as 
\begin{equation}
\rho  =  \frac{M\rho_{\mbox{eff}}(1+\tau+\sqrt{(1+\tau)^2+\frac{4\tau} { \rho_{\mbox{eff}}} } ) }{2t}
\label{eq:rhofrhoe}
\end{equation}

\section{Metrics}

The performance metric that we consider in this paper is the outage probability for a target rate. We evaluate the transmit power as a function of the number of antennas for a given coherence time with the outage $P_{out}$ constrained for a target rate $R$. We also find the least power that can achieve the target rate and outage by optimizing over the number of antennas.

We have found in (\ref{eq:cmodel2n}) an effective SNR under the assumption of worst case noise. We only deal with the model where power allocation is uniform among all the antennas and so the rate we calculate is not the channel capacity but the achievable rate. With this model we find a lower bound $\gamma$ on the achievable rate, in bits per second per channel use, from \cite{mimot} as

\begin{eqnarray}
\gamma & = &\zeta^{-1} \log{\det(I+\rho_{eff} \frac{ \bar{\Hm}\bar{\Hm}^{H} }{M})}
\label{eq:mimoc}
\end{eqnarray}

If the target data rate is represented by $R$, the outage probability $P_{out}$ is defined as the probability that the rate in a channel realization (mutual information), $\gamma$ is lower than $R$ the target rate.

\begin{equation}
P_{out} = P(\gamma < R)
\label{eq:effmodel}
\end{equation}

If $\mathcal{P}$ is the threshold outage probability required to maintain the quality of service (QoS), the performance metric has to satisfy the constraint $P_{out} \leq \mathcal{P}$. The transmission power is evaluated under two schemes, the fixed power scheme where transmission is always done at a constant power and the adaptive power scheme where transmission is done at the optimal power to achieve the target outage rate. Now, we discuss these two schemes in detail:

\subsection {Fixed power scheme}

The first scheme is applicable when transmission is always done with a constant power of $\rho_0$ and we shall refer to this as {\it fixed power scheme}. We calculate the $\rho_0$ as $\min(\rho  \mbox{, such as } P_{out}(R,\rho) \leq \mathcal{P})$ which is the lowest SNR that can achieve the given outage probability for a target rate in a MIMO system. Analyzing the behavior of $\rho_0$ with respect to changes in coherence time and the configuration of the MIMO system will be useful to determine how mobility affects the transmission power. 

\subsection {Adaptive power scheme}

The second scheme assumes that power control is implemented. In this model, the estimated channel state information (CSI) is sent back to the transmitting antennas through a feedback mechanism, which we assume to be instantaneous for simplicity. This assumption a a gross simplification in reality and especially when considering small coherence times, however as the qualitative results obtained will not very much different, we ignore the feedback load. The feedback in downlink MIMO is studied and optimized in \cite{feedb} and other related works and can be used to extend this paper to include its effects. Based on the feedback, the optimal power to achieve the target rate is used for transmission. If the power required is more than $\rho_0$ (The constant power defined in the previous sub-section), transmission is halted\footnote{This corresponds to a system where real time data processing is required like in voice or video calls where the rate is fixed and the packets are dropped if the power is insufficient as the transfer rate is fixed. For elastic services it is possible to lower the rate and continue communication with the maximum available power.}. As the power in this scheme is a function of the channel, for tractable measurements, the power calculated $\rho_{min}$ in this case is the average power over all possible channel realizations, and we refer to this model as {\it adaptive power scheme}. This model might be harder to implement due to the feedback mechanism required and larger Peak to average power ratio (PAPR). 

\begin{equation}
\rho_{min} = \mathbb{E}_{\Hm}(\rho_a(\Hm))
\label{rhoadaptive}
\end{equation}
where
\[\rho_a(\Hm) = \left\{ 
\begin{array}{l l}
  \rho(\Hm) & \quad \mbox{if $\rho(\Hm) \leq \rho_0$}\\
  0 & \quad \mbox{otherwise}\\ \end{array} \right. \]

\section{Theoretical analysis}

In this section we find expressions relating the outage rate $P_{out}$ and, the transmit power $\rho_0$ in the fixed power scheme and $\rho_{min}$ in the adaptive power scheme.

\subsection{MISO}

For the MISO system the optimal power allocation for the best outage probability has been conjectured in \cite{telatar} and later on proved in \cite{misoout}. Here we look at a MISO system with uniform power allocation. In this case, $\bar{\Hm}\bar{\Hm}^{H}$ reduces to a real number and $\det(I+\rho_{eff} \frac{ \bar{\Hm}\bar{\Hm}^{H} }{M})$ becomes just $1 + \sum_{i=1}^M\bar{h}_{ij}\bar{h}_{ij}^*$. Thus we can rewrite (\ref{eq:mimoc}) and (\ref{eq:effmodel}) as
\begin{eqnarray}
\rho_{eff} &=& \frac{M(\exp(R\zeta)-1)}{\sum_{i=1}^M\sum_{j=1}^N \bar{h}_{ij}\bar{h}_{ij}^* } \nonumber \\
&    =&  \frac{M(\exp(R\zeta)-1)}{\omega^2 }
\label{eq:minmisop}
\end{eqnarray}where $\sum_{i=1}^M\bar{h}_{ij}\bar{h}_{ij}^*$ is denoted as $\omega^2$. The distribution of $\omega^2$ is the Chi square distribution with 2M degrees of freedom, $\omega^2 \sim \chi^2(2M)$ \cite{chis}. If the maximum power $\rho$ is $\rho_0$, corresponding to a $\rho_{eff}$ of $\rho_{eff0}$ then, the outage is just the probability that $\rho_{eff0} \leq \frac{M(\exp(R)-1)}{{\bf Trace}(\bar{\Hm}\bar{\Hm}^{H}) }$. Defining $\Omega_0^2= \frac{M(\exp(R\zeta)-1)}{\rho_{eff0}}$, the outage can be calculated as
\begin{eqnarray}
P_{out} &=& \frac{\int_{0}^{\Omega_0}\omega^{M-1}\exp(-\frac{\omega}{2}) d\omega}{2^M \Gamma(M)}\nonumber \\
& = & \frac{ {\bf \it \gamma}(M,\frac{\Omega_0}{2})  }{\Gamma(M)}
\label{eq:misoouts}
\end{eqnarray}where $\Gamma$ is the Gamma function and ${\bf \it \gamma}$ is the lower incomplete Gamma function. The minimum average effective SNR, with which this outage is achieved, can be calculated from (\ref{rhoadaptive}), (\ref{eq:minmisop}) and (\ref{eq:misoouts}) as
\begin{eqnarray}
\rho_{\mbox{min-eff}} & = & \frac{M\exp(R\zeta-1)}{2^M \Gamma(M)}\int_{\Omega_0}^{\infty}\omega^{M-3}\exp(-\frac{\omega}{2}) d\omega \nonumber \\
& = & \frac{M\exp(R\zeta-1){\bf  \it \Gamma}(M,\frac{\Omega_0}{2}) }{\Gamma(M)}
\end{eqnarray}where ${\bf \it \Gamma}$ is the upper incomplete Gamma function. The actual average power in the adaptive power scheme can be obtained as $\rho_{min}=\rho(\rho_{\mbox{min-eff}})$ using (\ref{eq:rhofrhoe}).

An interesting observation can be made while comparing a single input single output (SISO) system and a MISO system with $M>1$. If we assume that the peak power can be arbitrarily large to obtain an arbitrarily small outage, then the SISO system in the adaptive power scheme will also consume an arbitrarily large power given by $\rho_{min} \propto (\lim_{\rho_0 \to \infty} {\bf Ei}(\frac{-1}{rho_0}) = \infty$), where {\bf Ei} is the exponential integral that approaches infinity in this limit. However the MISO system, with $M=2$ for instance, will only consume a finite average power $\rho_{min} \propto (\lim_{\rho_0 \to \infty} \exp(\frac{-1}{rho_0})=1)$. Outage in both cases is $\lim_{\rho_0 \to \infty} 1 - \exp(\frac{-1}{rho_0})=0$.

\subsection{SIMO}

Let us now look at a SIMO system with uniform power allocation. In this case, $\bar{\Hm}\bar{\Hm}^H$ is a matrix of rank one, so it has only one non-zero eigen value. This non-zero eigen value is also the trace and is given as $\sum_{j=1}^N\bar{h}_{ij}\bar{h}_{ij}^*$. $\det(I+\rho_{eff} \bar{\Hm}\bar{\Hm}^H )$ becomes just $1 + \sum_{j=1}^N\bar{h}_{ij}\bar{h}_{ij}^*$ as all the other eigen values are zero. Thus we can rewrite (\ref{eq:mimoc}) and (\ref{eq:effmodel}) as
\begin{eqnarray}
\rho_{eff} &=&  \frac{(\exp(R\zeta)-1)}{\omega_*^2}
\label{eq:minsimop}
\end{eqnarray}where $\sum_{j=1}^N\bar{h}_{ij}\bar{h}_{ij}^*$ is denoted as $\omega_*^2$. The distribution of $\omega_*^2$ is also the Chi square distribution with 2N degrees of freedom, , $\omega_*^2 \sim \chi^2(2N)$. If the maximum power $\rho$ is $\rho_0$, corresponding to a $\rho_{eff}$ of $\rho_{eff0}$ then, the outage is just the probability that $\rho_{eff0} \leq \frac{(\exp(R\zeta)-1)}{{\bf Trace}(\bar{\Hm}\bar{\Hm}^{H}) }$. Defining $\Omega_{0*}^2= \frac{(\exp(R\zeta)-1)}{\rho_{eff0}}$, the outage can be calculated as
\begin{eqnarray}
P_{out} &=& \frac{ {\bf \it \gamma}(N,\frac{\Omega_{0*}}{2} )  }{\Gamma(N)}
\label{eq:simoouts}
\end{eqnarray}where $\Gamma$ is the Gamma function and ${\bf \it \gamma}$ is the lower incomplete Gamma function. The minimum average effective SNR with which this outage can be achieved can be calculated from (\ref{rhoadaptive}), (\ref{eq:minsimop}) and (\ref{eq:simoouts}) as
\begin{eqnarray}
\rho_{\mbox{min-eff}} & = & \frac{\exp(R\zeta-1){\bf  \it \Gamma}(N,\frac{\Omega{0*}}{2}) }{\Gamma(N)}
\end{eqnarray}where ${\bf \it \Gamma}$ is the upper incomplete Gamma function. The actual average power in the adaptive power scheme can be obtained as $\rho_{min}=\rho(\rho_{\mbox{min-eff}})$ using (\ref{eq:rhofrhoe}).

\subsection{MIMO}

Calculating the outage probability in a general MIMO system is a tedious process and also it does not give an interpretable closed-form expression like in MISO or SIMO. Therefore, here we use results from the field of large random matrices to solve this problem and and show that the approximation holds even for a finite number of antennas.

\newtheorem{lem}{Lemma}
\begin{lem}

Given $\Hm \in \mathbb{C}^{N \times M}$ such that $h_{i,j} \sim C\mathcal{N}(0,1)$. As $M,N \to \infty$ such that $\lim_{N,M \to \infty} \frac{N}{M} = c $,
\begin{equation}
\log\det(I+\frac{\rho_{eff}}{M}(\Hm\Hm^H)) -M\mu  = \psi^o
\label{eq:mimormtbook}
\end{equation}
Where $\mu =[c\log(1+\rho_{eff}-\rho_{eff}\alpha)+\log(1+c\rho_{eff}-\rho_{eff}\alpha)-\alpha]$\\
and $\alpha = \frac{1}{2}[1+c+\rho_{eff}^{-1} - \sqrt{(1+c+\rho_{eff}^{-1})^2 -4c}]$ \\
Then, $\psi^o \xrightarrow{D} \psi $ (converges in distribution) where $\psi \sim \mathcal{N}(0,\sigma)$ with $\sigma=-\log(1-\frac{\alpha^2}{c})$. The proof is in \cite{rmtbook}.
\end{lem}

Lemma 1 gives us a simple and tractable expression to obtain a relation between the transmit power and the achievable rate. Additionally, we know how the rate converges in distribution allowing us to calculate the outage by transforming the complexity of working with $2MN$ random variables $\bar{\Hm}$ to a single random variable $\psi$. 
Now let us consider a MIMO system. From Lemma 1 we know that if the transmit power is $\rho_0$, the effective SNR can be found as $\rho_{eff0}$. If we define $\sigma_0=R\zeta - M\mu(\rho_{eff0})$, we have outage as
\begin{eqnarray}
P_{out} & \approx & P( \psi < R\zeta - M\mu_0)  \nonumber \\
  & = &\frac{ \int_{-\infty}^{R\zeta - M\mu_0 } \exp(\frac{-\psi^2}{2\sigma^2})d\psi}{\sqrt{2\pi\sigma^2}} \nonumber \\
  & = & 1 - Q(\frac{M\mu_0 - R\zeta }{\sigma})
\label{eq:mimormtout}
\end{eqnarray}where the $Q$ function is the tail probability of the standard normal distribution. This approximation and resulting calculations are verified by simulations in the next section. Consecutively, the minimum average SNR can be calculated from (\ref{rhoadaptive}), (\ref{eq:mimormtbook}) and (\ref{eq:mimormtout}) as
\begin{equation}
\rho_{\mbox{min}}= \frac{ \int_{M\mu_0 - R\zeta}^{\infty}\rho(\rho_{eff}(\psi))\exp(\frac{-\psi^2}{2\sigma^2})d\psi}{\sqrt{2\pi\sigma^2}}
\label{eq:mimominpow}
\end{equation}

\subsection{The inverse calculation}

In the previous sections we have detailed equations that express the outage probability as a function of $\rho_0$. However, our objective is to find $\rho_0$ for a given threshold outage probability. This is achieved following these series of steps. 

$\bullet$ Given an Outage probability threshold $P_0$. Since the $Q$-function is well known and invertable, we use the inverse $Q$-function to find $X=\frac{M\mu_0 - R\zeta }{\sigma}$.

$\bullet$ Now we use a linear search with $\rho$, M and $t$ as the parameters to calculate $\frac{M\mu_0 - R\zeta }{\sigma}$ and match it with the $X$ such that $\rho_0=min(\rho; \frac{M\mu_0 - R\zeta }{\sigma} \geq X)$. Where $\mu_0$ and $\sigma$ are evaluated as functions of $\rho_{eff}(\rho)$.

$\bullet$ Now using \ref{eq:mimominpow} we evaluate $\rho_{min}$.

The results of these calculations are illustrated in the following section with our numerical results.

\section{Numerical results}

All theoretical results obtained in the Section 4 are verified using Monte-Carlo simulations. The theoretical results are compared to the results from simulation by a linear search over $\rho$ that yields the desired $P_{out}$. Finally, we find $\{\rho_0, M, N, \tau \}$, such that $\rho_0 = \min(\rho_0(M, N, \tau) \mbox{, such as }M,N \in \mathbb{N}, \tau \geq 1)$ and $\{\rho_{min}, M^*, N^*, \tau^* \}$, such that $\rho_{min} = \min(\rho_{min}(M^*, N^*, \tau^*) \mbox{, such as }M^*,N^* \in \mathbb{N}, \tau^* \geq 1)$.

\subsection{MISO}

 \begin{figure}[h]
    \begin{center}
        \includegraphics[width=90mm]{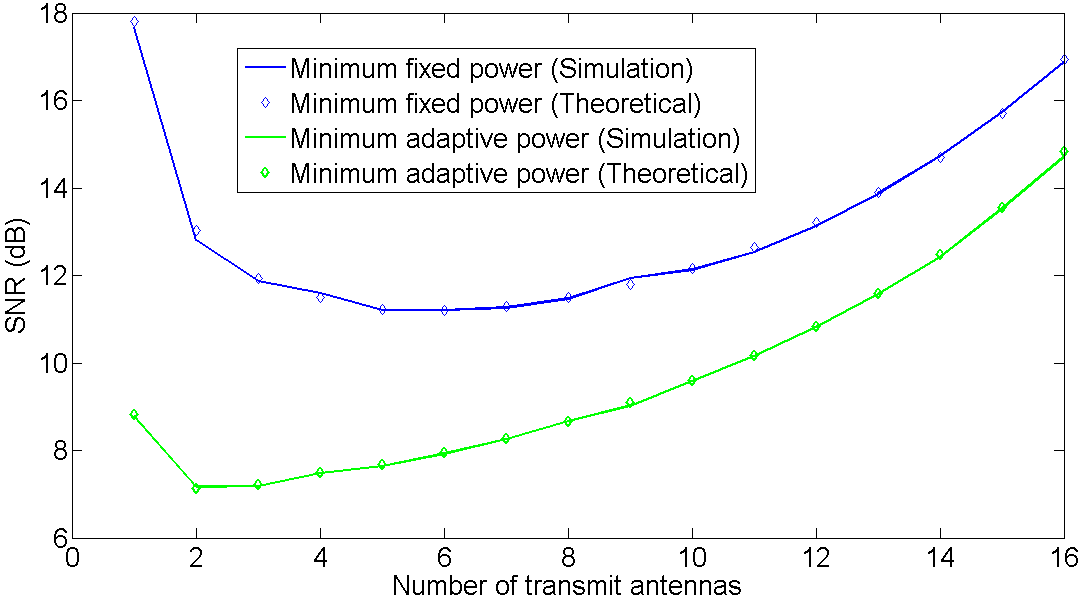}
    \end{center}
    \caption{ MISO system with $R = 1.44$ bps per hz, $\mathcal{P}= 5$\% and $T=25$ symbols. }
    \label{fig:misolowt}
\end{figure}

The SNR, $\rho_0$ in the fixed power scheme and $\rho_{min}$ in the adaptive power scheme for a MISO system, with a target rate of 1.44 bps, outage rate of 5\% and coherence time of 25 symbols, are plotted against $M=1,..16$ in Figure \ref{fig:misolowt}. It is clear from Figure \ref{fig:misolowt} that there is an optimal number of antennas for which the transmit power is minimized given a coherence time. As the number of antennas increases the gain from the additional degrees of freedom is lost due to the additional training time required. Another noteworthy fact is that the optimal number of antennas depends on the scheme of power transmission. The explanation for this is that the average SNR in the adaptive power scheme ($\rho_{min}$) increases as $M$ grows larger due to the channel hardening effect, while $\rho_0$ is still decreasing due to the gain from higher degrees of freedom. While $\rho_0$ takes its lowest value at $M=6$, $\rho_{min}$ is minimized at $M=2$. 

\subsection{SIMO}

\begin{figure}[h]
    \begin{center}
        \includegraphics[width=90mm]{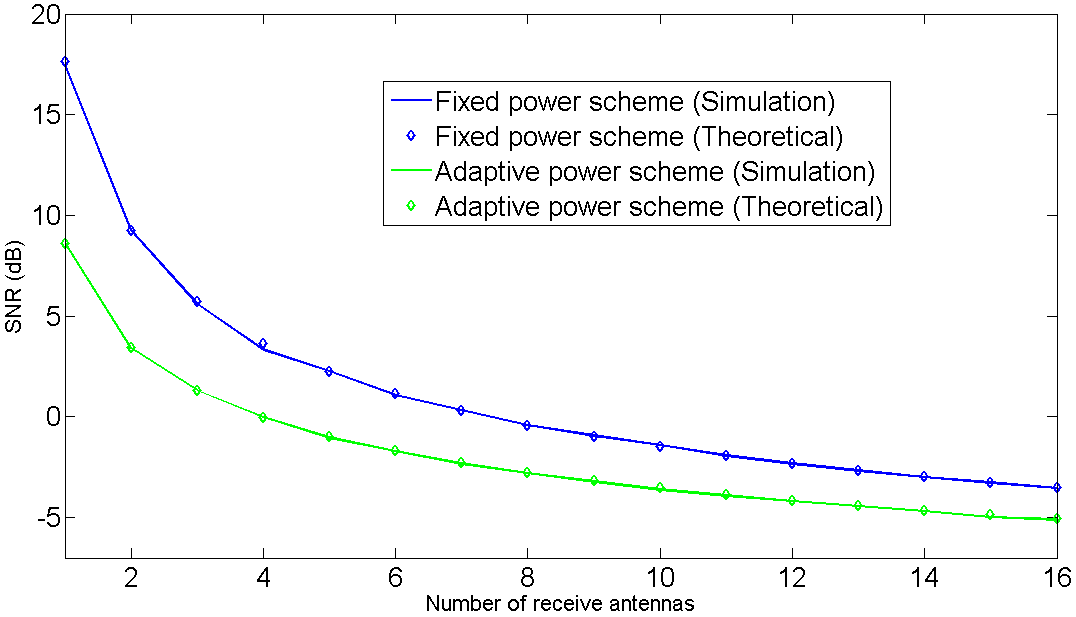}
    \end{center}
    \caption{ SIMO system with $R = 1.44$ bps per hz, $\mathcal{P}= 5$\% and $T=25$ symbols.}
    \label{fig:simolowt}
\end{figure}

We also consider a SIMO system with the same parameters as the MISO system. The SNR, $\rho_0$ in the fixed power scheme and $\rho_{min}$ in the adaptive power scheme are plotted against the number of antennas in Figure \ref{fig:simolowt}. Here we can see that the SNR is a monotonically decreasing function of the number of antennas. However, here we ignore the computational power required and there is no additional training required for the receiving antennas, so this result is expected. Thus, increasing the number of receive antennas always improves the achievable rate thus decreasing $\rho_0$ and $\rho_{min}$.

\subsection{MIMO}

 \begin{figure}[h]
    \begin{center}
        \includegraphics[width=90mm]{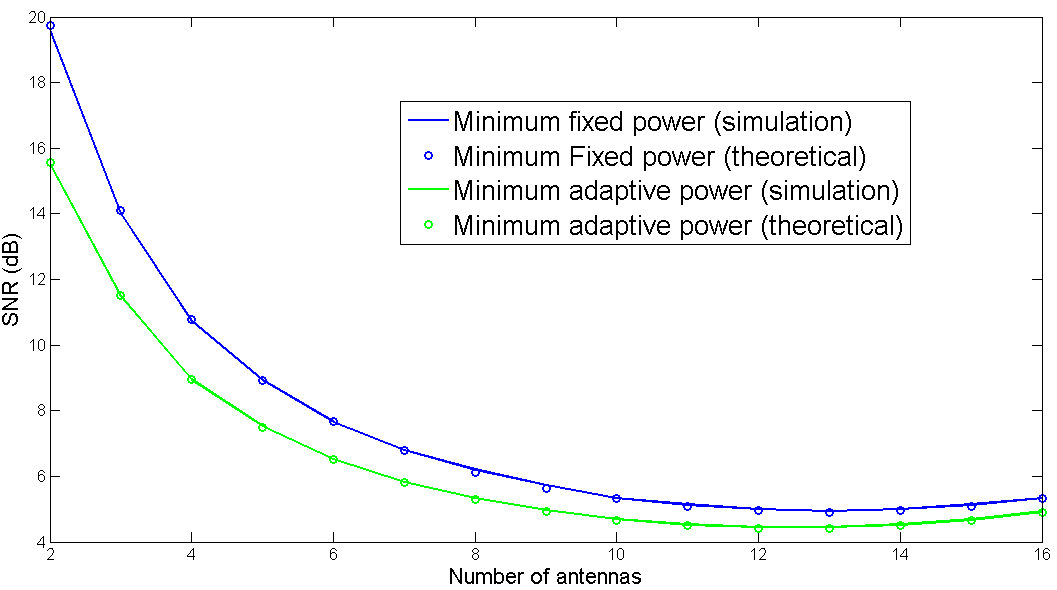}
    \end{center}
    \caption{\fontsize{6pt}{8pt} MIMO system with $R = 5.76$ bps per hz, $\mathcal{P}= 5$\%, $M=N$ and $T=25$ symbols.}
    \label{fig:mimolowt}
\end{figure}

\begin{figure}[h]
    \begin{center}
        \includegraphics[width=90mm]{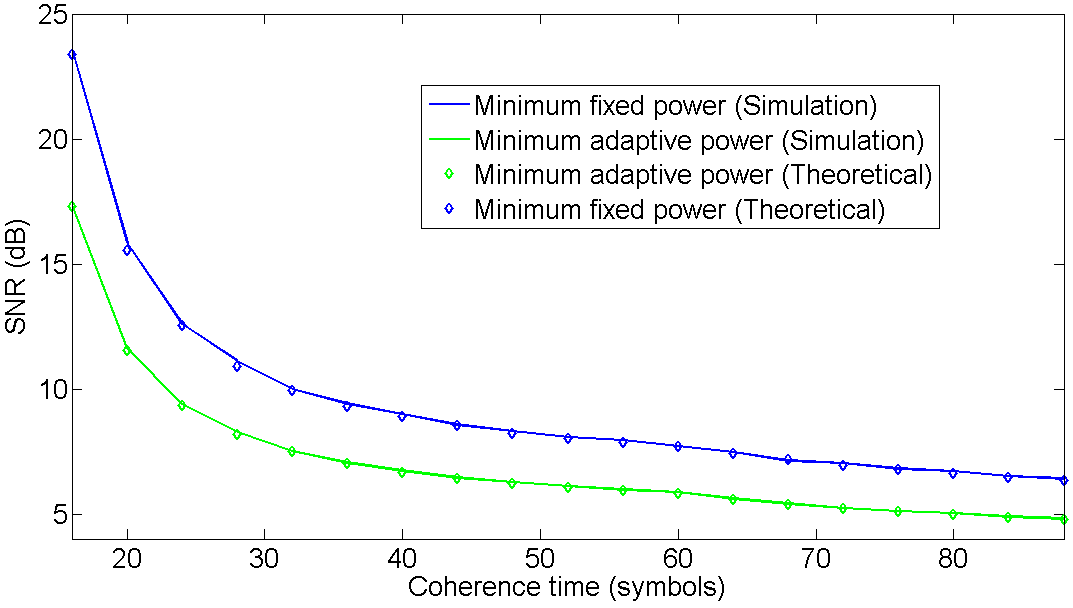}
    \end{center}
    \caption{MIMO system with $R = 5.76$ bps per hz, $\mathcal{P}= 5$\%, $M=8$ and $N=4$.}
    \label{fig:mimovt}
\end{figure}

The transmit power $\rho_0$ in fixed power scheme and $\rho_{min}$ in adaptive power scheme for a MIMO system with a target rate of 5.76 bps and outage probability of 5\% is plotted against the number of antennas ($N=M$) with $T=25$ in Figure \ref{fig:mimolowt}. We see that as the number of antennas increases the gain from the additional degrees of freedom and increased capacity is lost due to the additional training time required. The power in this case is minimized when $M=N=13$.

Figure \ref{fig:mimovt} plots the transmit power as a function of the coherence time for a MIMO system with $M=8$ and $N=4$ (target rate of 5.75 bps and outage of less than 5\%). We see that as the coherence time increases the transmit power decreases as expected. The slope of the curve also decreases as $T$ increases.

\subsection{Application}

Let us consider a MIMO system operating between a single BS and a user with a mobile terminal of four receiving antennas. The long term evolution (LTE) standards specify the symbol duration to be $66.7\mu$s \cite{lte}. Let us consider a carrier frequency of 3 Ghz. If we consider a highly mobile user with a speed of 100kmph, the coherence time in symbols is about 55. Using equation (\ref{eq:mimormtout}), we find the optimal number of transmit antennas and the training time for a data rate of 16bps per hertz (LTE spectral efficiency can be up to 15 bps per hertz) and outage probability of less than 1\%. Using a linear search on $M$ and $t$, we get the optimal number of antennas to be 8 and the training time to be 8 symbols corresponding to a minimum required SNR of 20 dB, while using 4 or 16 antennas would cost over 22 dB. If we consider a user with low mobility with a speed of about 10kmph, the coherence time is 550. The optimal number of antennas in this case would be 17 and the training time is 42 symbols. This corresponds to a minimum required SNR of about 16 dB while using just 4 antennas would cost over 20 dB.

\section{Conclusion}

In this paper, we have studied the impact of mobility on MISO, SIMO and MIMO systems. In order to have a tractable expression of the outage for a MIMO system we used recent results from the field of large random matrices. Simulations show these results to be tight even for two transmit and receive antennas. The quality of service is measured through the outage probability for a target rate in all cases and equations relating the outage probability to the transmit power were found. As a result, we see that given a coherence time, there is an optimal number of antennas for which the power required to transmit at a certain rate with a target outage probability is minimal. Studying a typical LTE system, we find that optimizing the number of antennas and training time can reduce power consumption up to 60\% and that using adaptive power control can further reduce the power by 60\%. Possible extensions of this work include the case of multi-user networks and studying the effects of the load due to feedback.

\section {Acknowledgements}

We would like to thank \emph{LSS, SUPELEC, The Alcatel Lucent Chair} and \emph{France Telecom} for providing the infrastructure and funding that helped in bringing this work to fruition. Additionally we would also like the thank J. Hoydis and V. Belmega for their insights and remarks that helped in editing and improving this work.



\end{document}